\documentclass[aoas,nameyear,seceqn,dvips]{arximspdf}
\usepackage{graphics}
\doi{10.1214/09-AOAS270}
\volume{3}
\issue{4}
\pubyear{2009}
\firstpage{1505}
\lastpage{1520}

\begin{document}
\begin{frontmatter}

\title{Using epidemic prevalence data to jointly estimate reproduction
and removal}
\runtitle{Jointly estimate reproduction and removal}

\begin{aug}
\author[A]{\fnms{Jan} \snm{van den Broek}\ead[label=e1]{j.vandenbroek@uu.nl}\corref{}}
and
\author[A]{\fnms{Hiroshi} \snm{Nishiura}\ead[label=e2]{h.nishiura@uu.nl}}
\runauthor{J. van den Broek and H. Nishiura}

\affiliation{Utrecht University}
\address[A]{Faculty of Veterinary Medicine\\
Utrecht University\\
The Netherlands\\
\printead{e1}\\
\phantom{E-mail:\ }\printead*{e2}} 
\end{aug}

\received{\smonth{1} \syear{2009}}
\revised{\smonth{6} \syear{2009}}

%
\begin{abstract}
This study proposes a nonhomogeneous birth--death model which captures
the dynamics of a directly transmitted infectious disease. Our model
accounts for an important aspect of observed epidemic data in which
only symptomatic infecteds are observed. The nonhomogeneous
birth--death process depends on survival distributions of reproduction
and removal, which jointly yield an estimate of the effective
reproduction number $R(t)$ as a function of epidemic time. We employ
the Burr distribution family for the survival functions and, as special
cases, proportional rate and accelerated event-time models are also
employed for the parameter estimation procedure. As an example, our
model is applied to an outbreak of avian influenza (H7N7) in the
Netherlands, 2003, confirming that the conditional estimate of $R(t)$
declined below unity for the first time on day 23 since the detection
of the index case.
\end{abstract}


\begin{keyword}
\kwd{Nonhomogeneous birth--death process}
\kwd{epidemic}
\kwd{double-binomial}
\kwd{Burr distribution}
\kwd{proportional rate model}
\kwd{accelerated event-time model}
\kwd{avian influenza}.
\end{keyword}

\end{frontmatter}

\section{Introduction}\label{sec1}
The data-generating process of an epidemic has special characteristics
to which one wants to pay particular attention when modeling these
data. First, the observed data of an infectious disease outbreak are
limited in the sense that the incidence---expressed as the number of
newly infected individuals as a function of time---is usually measured
by symptom onset of disease, which is sometimes further accompanied by
reporting delay. Thus, all of the observed cases in the reported data
represent those who experienced infection at some point in time in the
past. Second, epidemic data sets do not usually include information on
the number of susceptible individuals as a function of time, but solely
records infected (interpreted as symptomatic) individuals. It is
therefore unknown if susceptible individuals in the past are still
susceptible at a point of time. Third, the susceptible population is
usually not well defined at the beginning of an outbreak, and its size
may vary with time due to time-dependency in contact behavior and
public health countermeasures during the outbreak. In a veterinary
context, the countermeasures might include a transportation ban during
an infectious disease outbreak on animal farms. Control measures are
taken not only to reduce the number of contacts, but also to limit the
ability of infected individuals to generate secondary cases. For
instance, one may think of preemptive culling in the case of an
infectious disease outbreak on animal farms. Fourth, since the
infection is transmitted from individual to individual, observation of
an infected individual is not independent of observing other individuals.

These characteristics lead us to consider developing a method
which appropriately captures the dynamics of a directly transmitted
infectious disease by modeling the number of infected-and-detected
individuals, preferably in discrete time, in order to quantify the
reproduction of infected individuals in a nonhomogeneous manner. This
contrasts with other statistical models which measure the population of
susceptibles and model the force of infection at which these
susceptibles get infected.

As is usually assumed, one can think of a population in which
an epidemic of an infectious disease occurs\ as consisting of three
groups (or sub-populations) of individuals. The first is the
susceptible population which represents individuals who have not been
infected yet but may experience infection in the future. The second is
a population of infectious individuals which consists of those who have
been infected and are infectious to others. The last group consists of
removed or recovered individuals who are no longer infectious and may
be immune or are removed from the population. The simplest type of the
model which describes the transmission dynamics over time is referred
to as an SIR (Susceptible-Infected-Recovered) model [\citet
{diekman-heesterbeek}]. Since the present study will concentrate on the
number of infected, here we consider their dynamics alone. Letting
$x(t)$ and $y(t)$ be the susceptible and infectious fractions of the
population at time $t$, respectively, the derivative of $y(t)$ is
expressed as
\begin{equation}\label{sir}
\frac{\mathrm{d}} {\mathrm{d}t} y(t)=\beta(t) y(t) x(t)- \mu(t) y(t).
\end{equation}
Note that the transmission rate $\beta(t)$ and the removal rate $\mu
(t)$ may depend on time. If one rewrites the product of the
transmission rate $\beta(t)$ and susceptibles $x(t)$ as $\lambda(t)$
($=\beta(t)x(t)$), then the equation is rewritten as
\begin{equation}\label{det-bd}
\frac{\mathrm{d}} {\mathrm{d}t} y(t)=\lambda(t) y(t) - \mu(t) y(t),
\end{equation}
which is the equation of the deterministic nonhomogeneous birth--death
process. The function $\lambda(t)$ has been referred to as the
reproductive power and can be interpreted as the rate at which a single
infected individual is able to generate secondary cases [\citet
{kendall}]. In other words, $\lambda(t)$ is the rate at which an
infected individual is able to reproduce itself. The so-called death
rate $\mu(t)$ in a birth--death process is interpreted as the rate at
which an infected individual is removed from the sub-population of
infected individuals. It should be noted that equation (\ref{det-bd})
relaxed the definition of $y(t)$ compared to that in (\ref{sir}).
Namely, whereas $y(t)$ in (\ref{sir}) has to be \textit{infectious} to
others, we can instead regard $y(t)$ in (\ref{det-bd}) as \textit{infected-and-detected} individuals (i.e., regardless of infectiousness).

One of the advantages of using this simple equation is that the
population of susceptibles is allowed to vary over time. Therefore, the
reproductive power varies over time due to two different reasons: (1)
the population of susceptibles $x(t)$ varies as a function of time and
(2) the transmission rate $\beta(t)$ is nonhomogeneous over time. In
addition, the removal rate $\mu(t)$ is allowed to vary with time.

It can be an advantage to model infection process
stochastically, because one can explicitly define the probability of
transmission, rather than deterministically stating if the transmission
happens [\citet{andersson}]. A stochastic model can describe not only
the quantitative patterns of observation with time-dependent expected
values, but also offer standard errors of the parameters without making
adhoc distributional assumptions. More importantly, the likelihood
function can be explicitly derived, which will be useful for
statistical inference of parameters and critical assessments of the
modeling method. Moreover, such a stochastic process can model the
number of infected over time as being dependent.

The present study aims to develop a stochastic model which is
based on a nonhomogeneous birth--death process. The model is applied to
an observed data set of infected-and-detected (but not yet removed)
cases, permitting reasonable assessment of the time course of an
epidemic. In Section \ref{sec2} the stochastic version of the nonhomogeneous
birth--death model is comprehensively described. A novel analytical
solution of the model is obtained with the use of a general Lagrange
transformation derivation of which has not been explicitly discussed to
date. In Section \ref{sec3} we discuss a conditional discrete-time fitting
method. Although a similar conditional fitting procedure was employed
in a recent study [\citet{vandenbroek}], the present study is the first
to apply the technique to a model where both birth and death rates are
nonhomogeneous. Depending on the model and the data given, the number
of parameters to be estimated can be large and it might be difficult to
get stable estimates. Besides, a relationship between the reproductive
power and the removal rate might exist. Therefore, more restricted
models which employ this relationship are discussed in Section \ref{sec4}. In
Section \ref{sec5} our model is applied to an observed epidemic data set of
avian influenza A (H7N7) in the Netherlands, 2003.

\section{The nonhomogeneous birth--death model}\label{sec2}
The stochastic differential equations for a nonhomogeneous birth--death
process, with $Y(t)$ and $y_0$ being the number of
infected-and-detected at time $t$ and the initial number of
infected-and-detected at time $0$, respectively, are as follows:
\begin{eqnarray*}
\frac{\mathrm{d}}{\mathrm{d}t} p_y(t)&=&\lambda(t)(y-1)p_{y-1}(t)+\mu
(t)(y+1) p_{y+1}(t)- \bigl( \lambda(t)+\mu(t) \bigr) y p_y(t),\\
\frac{\mathrm{d}}{\mathrm{d}t} p_0(t)&=& \mu(t) p_1(t),
\end{eqnarray*}
where $\lambda(t)$ denotes the reproductive power, $\mu(t)$ the death
rate and $\Pr(Y(t)=y)=p_y(t)$ the probability that the number of
infected-and-detected individuals at time $t$ is $y$. It should be
noted that here we consider $Y(t)$ as the number of
infected-and-detected individuals which represents the observed
elements of the data and is irrelevant to infectiousness. If we take
the probability generating function of the probabilities $p$, we can
derive a partial differential equation for this fraction by multiplying
the above differential equations with $z^y$, summing the result, and
taking the derivative. The analytical solution of the partial
differential equation is [\citet{kendall}]
\[
\Phi(y,u)= \biggl[\frac{\theta-(1-\pi-\theta)u}{1-\pi u} \biggr]^{y_0}.
\]
If we write $\rho(t)=\int_0^t [\mu(\tau)-\lambda(\tau) ]\,\mathrm{d}\tau
=\log[\frac{S_{\lambda}(t)}{S_{\mu}(t)} ]$ and
\begin{equation}\label{gam}
\gamma(t)=\int_0^t e^{\rho(\tau)}\lambda(\tau)\,\mathrm{d}\tau=-\int
_0^t\frac{\mathrm{d}S_{\lambda}(\tau)}{S_{\mu}(\tau)},
\end{equation}
we get
\begin{eqnarray}
\label{thea}
\theta&=&1-\frac{e^{-\rho(t)}}{1+\gamma(t)e^{-\rho(t)}}=1-\frac{S_{\mu
}(t)}{S_{\lambda}(t)+\gamma(t)S_{\mu}(t)},
\\
\label{piet}
\pi&=&1-\frac{1}{1+\gamma(t)e^{-\rho(t)}}=1-\frac{S_{\lambda
}(t)}{S_{\lambda}(t)+\gamma(t)S_{\mu}(t)}.
\end{eqnarray}
It should be noted that $S_{\lambda}(t)=e^{-\int_0^t \lambda(\tau)
\,\mathrm{d}\tau}$ is the \textit{reproduction} survival function and $S_{\mu
}(t)=e^{-\int_0^t \mu(\tau) \,\mathrm{d}\tau}$ is the \textit{removal}
survival function.

To obtain the probability distribution from $\Phi(y,u)$, the
general Lagrange transformation is useful (the details of which can be
found elsewhere\break [\citet{consul}]). First, let $\Phi(y,u)= [\theta
+(1-\theta)\frac{(1-\pi)u}{1-\pi u} ]^{y_0}= [\theta+(1-\theta)\psi(u)
]^{y_0}$, where $\psi(u)$ is the probability generating function (pgf)
of the geometric distribution. Second, let $g(z)=1-\pi+\pi z$, the pgf
of a Bernoulli distribution. Numerically, the smallest root of the
transformation $z=ug(z)$ defines a pgf $z=\psi(u)=\frac{(1-\pi)u}{1-\pi
u}$ [\citet{consul}]. Third, additionally considering $f(z)=(\theta
+(1-\theta)z)^{y_0}$, the pgf of the discrete general Lagrange
probability distribution under the Lagrange transform $z=ug(z)$ is
given by $f(z)=f(\psi(u))= [\theta+(1-\theta)\frac{(1-\pi)u}{1-\pi u}
]^{y_0}$ and, moreover, the probability mass function is a special case
of the double-binomial distribution [\citet{consul}, pages 22--27], which
can be referred to as the Bernoulli-binomial Lagrangian distribution in
the terminology of [\citet{johnson}]
\begin{eqnarray}\label{dbinom}
P\bigl(Y(t)=0\bigr)&=& \theta^{y_0}, \nonumber\\
P\bigl(Y(t)=y\bigr)&=& \frac{y_0}{y}\theta^{y_0}\pi^{y}\sum_{k=0}^{\min
(y-1,y_0-1)} \pmatrix{y _0-1 \cr k}\pmatrix{y \cr y-k-1}\nonumber\\[-8pt]\\[-8pt]
&&\hphantom{\frac{y_0}{y}\theta^{y_0}\pi^{y}\sum_{k=0}^{\min
(y-1,y_0-1)}}
{}\times \biggl[ \frac{(1-\pi
)(1-\theta)}{\pi\theta} \biggr]^{k+1}, \nonumber\\
& &\hspace*{223pt} y\geq1,\hspace*{-223pt}\nonumber
\end{eqnarray}
where $\theta$ and $\pi$ are defined in equations (\ref{thea}) and (\ref
{piet}). It should be noted that $g(z)$ and $f(z)$ are pgf's and, thus,
the necessary conditions for Lagrange transformation are satisfied. To
the best of our knowledge, detailed derivation of equation
(\ref{dbinom}) has never been discussed in the context of the nonhomogeneous
birth--death model (see Section \ref{sec6}).

The expectation and the variance of (\ref{dbinom}) are
\begin{eqnarray*}
E(y)&=&y_0\frac{1-\theta}{1-\pi} \\
&=&y_0\frac{S_{\mu}(t)}{S_{\lambda}(t)}, \\
\operatorname{var}(y)&=&y_0\frac{(1-\theta)[\theta+\pi(1-\theta-\pi)]}{(1-\pi)^3} \\
&=&y_0\frac{S_\mu(t)}{S_\lambda(t)} \biggl[1+(2\gamma-1)\frac{S_\mu
(t)}{S_\lambda(t)} \biggr] \\
&=&y_0R(t) [1+(2\gamma-1)R(t) ]
\end{eqnarray*}
by using (\ref{thea}) and (\ref{piet}).

The expected value has two interpretations. The first part is the
predicted number of infected individuals at time $t_0$ who survived
removal [i.e., $y_0S_{\mu}(t)$]. The second part, $\frac{1}{S_{\lambda
}(t)}$, measures the rate at which a nonremoved infected individual
reproduces itself. This is similar to an interpretation of a
nonhomogeneous birth process [\citet{vandenbroek}]; the difference of
the present study from the previous nonhomogeneous birth process is
that in the present setup only the predicted nonremoved
infected-and-detected individuals reproduce. Second, the ratio of the
rates $\frac{S_{\mu}(t)}{S_{\lambda}(t)}$ is the net reproduction ratio
with which an infected individual reproduces itself, which is
interpreted as the effective reproduction number $R(t)$ as a function
of epidemic time $t$. $R(t)$ in the present study can be regarded as
the average number of secondary cases generated by a single primary
case at time $t$. That is, our $R(t)$ is an instantaneous measure of
secondary transmissions occurring at time $t$, whose definition is
equivalent to the period total fertility rate in mathematical
demography [\citet{nishiura}]. If $R(t) < 1$, it suggests that the
epidemic is in decline and may be regarded as being ``under control'' at
time $t$ [vice versa, if $R(t) > 1$]. It should be noted that the
expected value of (\ref{dbinom}) is equivalent to an analytical
solution of the deterministic version of a nonhomogeneous birth--death
process (\ref{det-bd}).

The term $\gamma(t)$ in the formula for the variance represents
the dependence between the birth and death rate as can be seen from
(\ref{gam}). As it is clear from the analytical expression for the
variance, the variance becomes large if the probability of nonremoval
is large for an infected individual, if the probability of reproduction
is large, or both. This matches intuitive sense. In addition, the
variance can be regarded as a type of negative binomial variance, if we
rewrite it as $\operatorname{var}(y)=E(y)[1+\frac{2\gamma-1}{y_0}E(y)]$.

\section{Fitting the model}\label{sec3}
We have shown how the epidemic data can be generated by a stochastic
nonhomogeneous birth--death process. Nevertheless, the observed data
are, in reality, just one sample path of all the possible sample paths
that can arise from such an epidemic process. Considering further that
the number of infected-and-detected and individuals at a certain point
in time $t$ depends on the number of infected-and-detected individuals
at some time point before $t$, our model is fitted to the data by
conditioning on the transmission dynamics which happened before $t$
[\citet{becker1}, \citet{becker2}]. Moreover, as we briefly discussed in the
\hyperref[sec1]{Introduction}, another important point in practice is the discrete
nature of the time points of observation, that is, say, $t_j$,
$j=0,1,\ldots,n$, where the time unit might typically be days or weeks.
Therefore, the number of infecteds at time $t_j$ is modeled
conditionally on the number of infecteds by time $t_{j-1}$.

Since the probability mass function (\ref{dbinom}) is a
conditional probability mass function, conditioning being on the number
of infecteds at $t_0$, this can be effectively used as the conditional
model for the number of infecteds at time $t_j$, given the number of
infecteds at time $t_{j-1}$. For this reason, the survival
distributions $S_{\lambda}(t)$ and $S_{\mu}(t)$ and, of course, $\gamma
(t)$, are also conditioned on the past. Let $T_{\lambda}$ and $T_{\mu}$
be the stochastic variables that measure the reproduction time and the
removal time, respectively. The conditional survival probability for
the reproduction time is
\begin{eqnarray*}
P(T_{\lambda} > t_j | T_{\lambda}> t_{j-1})&=&\frac{S_{\lambda
}(t_j)}{S_{\lambda}(t_{j-1})} \\
&=&1-\frac{S_{\lambda}(t_{j-1})-S_{\lambda}(t_j)}{S_{\lambda}(t_{j-1})}
\\
&=&1-P\bigl(T_{\lambda} \in(t_{j-1},t_j] |T_{\lambda}>t_{j-1}\bigr) \\
&=&1-h_{\lambda}(t_{j-1}),
\end{eqnarray*}
where $h_{\lambda}(t_{j-1})$ is interpreted as the discrete
reproductive power. Similarly, the conditional survival probability for
the removal time is given by
\[
P(T_{\mu} > t_j | T_{\mu}> t_{j-1})=1-h_{\mu}(t_{j-1}),
\]
where $h_{\mu}(t_{j-1})$ is the discrete removal hazard.

The discrete conditional version of (\ref{gam}) in the time
interval ($t_{j-1}$, $t_j$] is
\[
\frac{P(T_{\lambda} \in(t_{j-1},t_j] |T_{\lambda}>t_{j-1})}{P(T_{\mu}
> t_j | T_{\mu}> t_{j-1})}=\frac{h_{\lambda}(t_{j-1})}{1-h_{\mu}(t_{j-1})}.
\]
When these conditional discrete measurements are considered, $\theta$
and $\pi$ correspond to $h_{\mu}(t_{j-1})$ and $h_{\lambda}(t_{j-1})$,
respectively. Using these, the conditional probability for (\ref
{dbinom}) is expressed as
\[
P\bigl(Y(t_j)=0|Y(t_{j-1})=y_{t_{j-1}}\bigr)=h_{\mu}(t_{j-1})^{y_0},\vspace*{-6pt}
\]
\begin{eqnarray*}
&&P\bigl(Y(t_j)=y_{t_j}|Y(t_{j-1})=y_{t_{j-1}}\bigr)\\
&&\qquad =\frac
{y_{t_{j-1}}}{y_{tj}}h_{\mu}(t_{j-1})^{y_{t{j-1}}}h_{\lambda
}(t_{j-1})^{y_{t_j}}
\\
&&\quad\qquad{}\times\sum_{k=0}^{\min(y_{t_j}-1,y_{t_{j-1}}-1)} \pmatrix{y_{t_{j-1}}-1
\cr k}\pmatrix{y_{t_j} \cr y_{t_j}-k-1}\\
&&\hphantom{\quad\qquad{}\times\sum_{k=0}^{\min(y_{t_j}-1,y_{t_{j-1}}-1)}}
{}\times \biggl[ \frac{(1-h_{\lambda
}(t_{j-1}))(1-h_{\mu}(t_{j-1}))}{h_{\lambda}(t_{j-1})h_{\mu}(t_{j-1})}
\biggr]^{k+1},\qquad y\geq1.
\end{eqnarray*}
The expected value of this probability is $y_{t_{j-1}}\frac{1-h_{\mu
}(t_{j-1})}{1-h_{\lambda}(t_{j-1})}$, which is referred to as the
sample path profile [\citet{lindsey}]. The corresponding conditional
measurement of the effective reproduction number is $R(t_{j-1})=\frac
{1-h_\mu(t_{j-1})}{1-h_\lambda(t_{j-1})}, $ with $h_\mu(t_{j-1})=1-\frac
{S_\mu(t_j)}{S_\mu(t_{j-1})}$ and $h_\lambda(t_{j-1})=1-\frac{S_\lambda
(t_j)}{S_\lambda(t_{j-1})}$.

Let $\Delta$ be the vector of parameters from the survival
distributions (see Section~\ref{sec4}). The log-likelihood function is then
\[
l(\Delta)=\sum_{i=1}^n\log
\bigl[P\bigl(Y(t_j)=y_{t_j}|Y(t_{j-1})=y_{t_{j-1}},\Delta\bigr) \bigr].
\]
Note that the likelihood is evaluated only for $y_{t_j}>0$, because
zero prevalence is the absorbing state of the process and this state is
not observable in reality.

 The log-likelihood can be maximized using an optimization
procedure, such as the Nelder--Mead method to find the maximum
likelihood estimates. In our exercise, the software system R [\citet{r}]
is used. The information matrix is used to find the standard errors of
the parameters, and we use Akaike's information criterion (AIC) to
compare model fits.

\section{The Burr distribution and its special cases}\label{sec4}
To choose a particular form for the survival functions, one might take
the early phase of the outbreak into account. The mean of the
probability mass function (\ref{dbinom}) depends on these survival
functions and is the same as the solution of (\ref{det-bd}). Since (\ref
{det-bd}) can be derived from the SIR model, one might look at the
deterministic SIR-model to decide the parametric form of the survival
function. In the early phase of the outbreak a deterministic SIR-model
can be well approximated by a deterministic SI-model since in that
phase the number of removals is limited. The dynamic equations for this
SI-model hold for the fraction of susceptibles and for the fraction of
infected and since the fraction of susceptibles at a point of time is
the same as the fraction of individuals with infection time larger than
that time point, the dynamic equations should also hold for the
survival function.
The Burr family of distribution functions has this precise property
[\citet{vandenbroek}].

When detection of a symptomatic infected individual occurs,
he/she usually will be removed immediately. Thus, it is reasonable to
assume that the reproductive power and the removal rate have a similar
structure, and follow a similar survival function.

The most well-known and useful distribution from the Burr
family is the Burr XII, or Singh--Maddala distribution, which in the
literature is sometimes referred to simply as the Burr distribution.
The survival function is given by
\[
S(t)= \biggl[ 1+ \biggl(\frac{t}{b} \biggr)^a \biggr]^{-q},\qquad t>0, a,b,q>0.
\]
The right tail is governed by the parameters $a$ and $q$, the left tail
by $a$, and $b$ is the scale parameter [\citet{kleiber-kotz}, page 198].
To reduce the number of parameters to be estimated, one can consider
three special cases of the Burr distribution [\citet{kleiber-kotz}]:
\begin{enumerate}
\item The logistic form is obtained for $q=1,$ giving the log-logistic
or the Fisk distribution.
\item For $a=1$, the Burr distribution is reduced to the Lomax (Pareto
type II) distribution.
\item The case $a=q$ is also known as the para-logistic distribution.
\end{enumerate}
The Weibull distribution and the Pareto distribution are limiting cases
of the Burr distribution [Shao (\citeyear{shao})]. An interesting way to arrive at
the Burr distribution is to assume that the times follow a Weibull
distribution, the scale parameter of which follows an inverse
generalized gamma distribution [\citet{kleiber-kotz}].

As another way to reduce the number of model parameters, and to
find a further relationship between the reproductive power and the
removal rate, one may rewrite the Burr distribution as a proportional
rate or an accelerated event-time distribution (just as in the case of
the more famous Weibull distribution). Suppose that the survival
function for the reproduction time is a Burr distribution with
parameters $a, b_1$ and $q$ and suppose that the survival function of
the removal time is also a Burr distribution with parameters $a, b_2$
and $q$. If we replace $b_2$ by $db_1$ (where $d$ is a constant), we get
\[
S_{\mu}(t)= \biggl[ 1+ \biggl(\frac{t}{b_2} \biggr)^a \biggr]^{-q}= \biggl[ 1+ \biggl(\frac{t}{db_1} \biggr)^a
\biggr]^{-q}= \biggl[ 1+ \biggl(\frac{t^{\prime}}{b_1} \biggr)^a \biggr]^{-q}=S_{\lambda}(t^{\prime}).
\]
Therefore, the survival distribution for the removal times is exactly
the same as that for the reproduction time, except that the removal
time is interpreted as accelerated reproduction time.

To employ the proportional rate model, let the survival
distribution for the reproduction time be a Burr distribution with
parameters $a, b$ and $q_1$, and suppose that the survival distribution
of the removal times is also a Burr distribution with parameters $a, b$
and $q_2$. If we replace $q_2$ by $cq_1$ (where $c$ is a constant), we get
\[
S_{\mu}(t)= \biggl[ 1+ \biggl(\frac{t}{b} \biggr)^a \biggr]^{-q_2}= \biggl[ 1+ \biggl(\frac{t}{b} \biggr)^a
\biggr]^{-cq_1}= \biggl\{ \biggl[ 1+ \biggl(\frac{t}{b} \biggr)^a \biggr]^{-q_1} \biggr\}^c= \{S_{\lambda}(t) \}^c,
\]
indicating that the rates are proportional; that is, the rate at which
an infected-and-detected is removed is proportional to the rate at
which an infected-and-detected individual reproduces. Of course, the
Burr distribution can also be written as both an accelerated event-time
distribution and as a proportional rate distribution, that is, $S_{\mu
}(t)= [S_{\lambda}(t^{\prime}) ]^c $.

All the distributions described above can be written in
accelerated event-time form and in proportional hazard form, except for
log-logistic distribution which has only an accelerated event-time
form. In the next section we fit all of those models to epidemic data
of avian influenza A (H7N7) in the Netherlands, 2003.

\section{Dutch avian influenza A (H7N7) epidemic in 2003}\label{sec5}
Here we show an example of our model application to an observed data
set. An epidemic of avian influenza A (H7N7) virus started on February
28, 2003, in the Gelderse Vallei in the Netherlands. In total, 239
flocks experienced infection with known detection date. Control
measures taken include movement restrictions, stamping out of infected
flocks, and preemptive culling of flocks in the neighborhood of
infected flocks. As a result, 1255 commercial flocks and 17,421 flocks
of smallholders had to be depopulated, and approximately 25.6 million
animals were killed. The virus was also transmitted to humans who had
been in close contact with the infected chickens, resulting in one
human death. Further details can be found elsewhere [\citet{stegeman1}].

We examine transmission and detection events between flocks. We
regard the detection date of a case (i.e., infected individual) as the
date at which there were first signs of infection in a flock. In other
words, the detection date of an infected individual is regarded as the
birth date in our model. Therefore, the birth date is not the date of
infection but the date at which an infected farm is detected, which is
used as a surrogate. Moreover, the date of depopulation is regarded as
the death date. Consequently, the Dutch data consist of the prevalence
of infected-and-detected (but not yet removed) flocks on each epidemic date.

\begin{figure}

\includegraphics{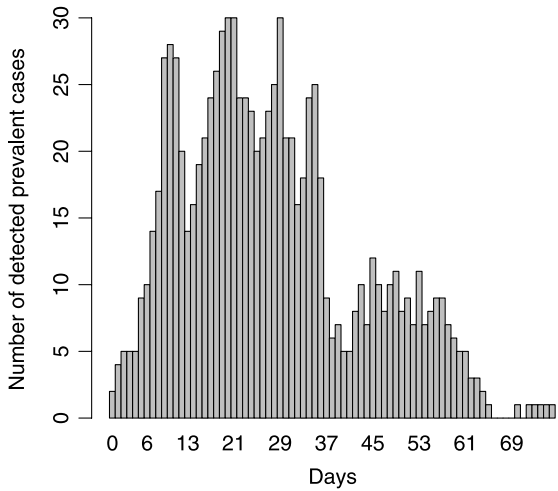}

\caption{\textup{Temporal distribution of the prevalent cases
of avian influenza A (H7N7) epidemic in the Netherlands, 2003}. The
index case was reported on February 28, 2003 (and the date is defined
as day 0). The prevalent cases represent those who have been infected
and detected but have not been depopulated yet at day $t$, which
correspond to the expected value of $y_t$ in Section \protect\ref{sec3}.
See Stegeman et al. \protect\citeyear{stegeman1} for further information.}\label{fig1}
\end{figure}

Figure \ref{fig1} shows the temporal distribution of the
prevalent cases (representing those who were born and have not been
removed yet). As can be seen, the right tail contains gaps and the
center of the distribution is not well determined. Usually these make
it difficult to fit simple models to the data.

\begin{table}[b]
\tabcolsep=0pt
\caption{Akaike's information criterion (AIC) for different models}\label{tab1}
\begin{tabular*}{\textwidth}{@{\extracolsep{\fill}}lcccc@{}}
\hline
&\textbf{Full model}&\textbf{Acc. event time\tabnoteref[1]{table11}} &\textbf{Prop. rate\tabnoteref[2]{table12}} &\textbf{Both acc. event and prop. rate} \\
\hline
Burr&342.4&345.90&346.4&342.7 \\
Log-logistic&371.3&369.8&--&-- \\
Lomax&341.9&344.8&344.9&same as full model \\
Para-logistic&351.8&354.0&357.5&same as full model \\
\hline
\end{tabular*}
\tabnotetext[1]{table11}{Accelerated event-time model.}
\tabnotetext[2]{table12}{Proportional rate model.}
\end{table}

\begin{figure}[t]

\includegraphics{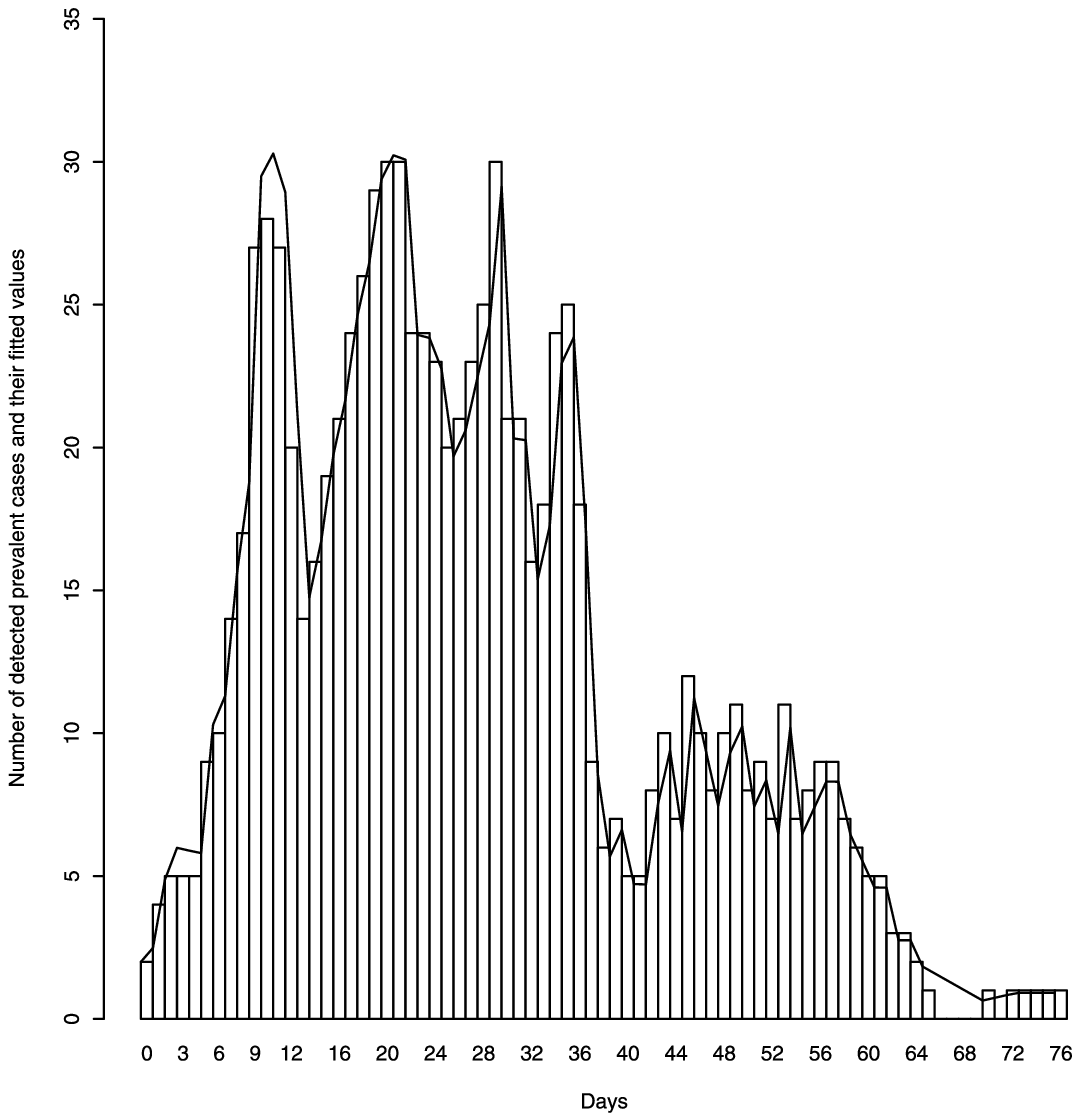}

\caption{\textup{Comparison of the observed numbers and the
predicted values from the conditional model of prevalent cases of avian
influenza A (H7N7) epidemic in the Netherlands, 2003}. The index case
was reported on February 28, 2003 (and the date is defined as day 0).
Observed data (bars) is compared with the predicted number of cases
(solid line) based on the full Lomax model. It should be noted that the
expectation of prevalence $y_{t_j}$ is conditioned on $y_{t_{j-1}}$.}\label{fig2}
\end{figure}

The Burr model with three parameters for birth rate and three
for death rate was fitted to the observed data. We refer to it as the
\textit{full} Burr model. To objectively show how this model fits to the
data better than other types of models, we compared its likelihood with
that of the full inverse Burr (Burr III or Dagum) model. The inverse
Burr model is also known as a flexible distribution from the Burr
family and can be viewed as a generalized gamma with a scale parameter
that follows an inverse Weibull distribution [\citet{kleiber-kotz}]. The
inverse Burr yielded an AIC of $363.7$, while the full Burr model
yielded 342.4. We thus examined the full Burr model (and its special
cases) for further analyses. The AICs for these different models are in
Table \ref{tab1}. The full Lomax model gave the best fit, although the
difference in AIC with the full Burr distribution was not particularly
large. In other words, the information criterium suggest that both the
death rate and the reproductive power may be proportional and that the
removal time may be accelerated reproduction time in the observed data
set. Figure \ref{fig2} visually confirms a good fit of the full Lomax
model to the observed number of infected-and-detected individuals based
on the conditional model in discrete time. The model seems to have well
captured the observation, because fitting prevalence $y_{t_j}$ to the
data is conditioned on $y_{t_{j-1}}$. It should be noted that the
predicted values in Figure \ref{fig2} reflect a qualitative pattern of
the observed data always one or two steps late, which is a general
tendency of conditional fit.

\begin{table}
\tablewidth=172pt
\caption{Parameter estimates for the Lomax distribution}\label{tab2}
\begin{tabular*}{172pt}{@{\extracolsep{\fill}}lcc@{}}
\hline
\textbf{Parameter}&\textbf{Estimate}&\textbf{St. error} \\
\hline
$\ln(b_1)$&3.235&0.5998 \\
$\ln(q_1)$&2.712&0.3611 \\
$\ln(b_2)$&4.987&1.0396 \\
$\ln(q_2)$&3.980&0.8480 \\
\hline
\end{tabular*}
\end{table}

\begin{figure}[b]

\includegraphics{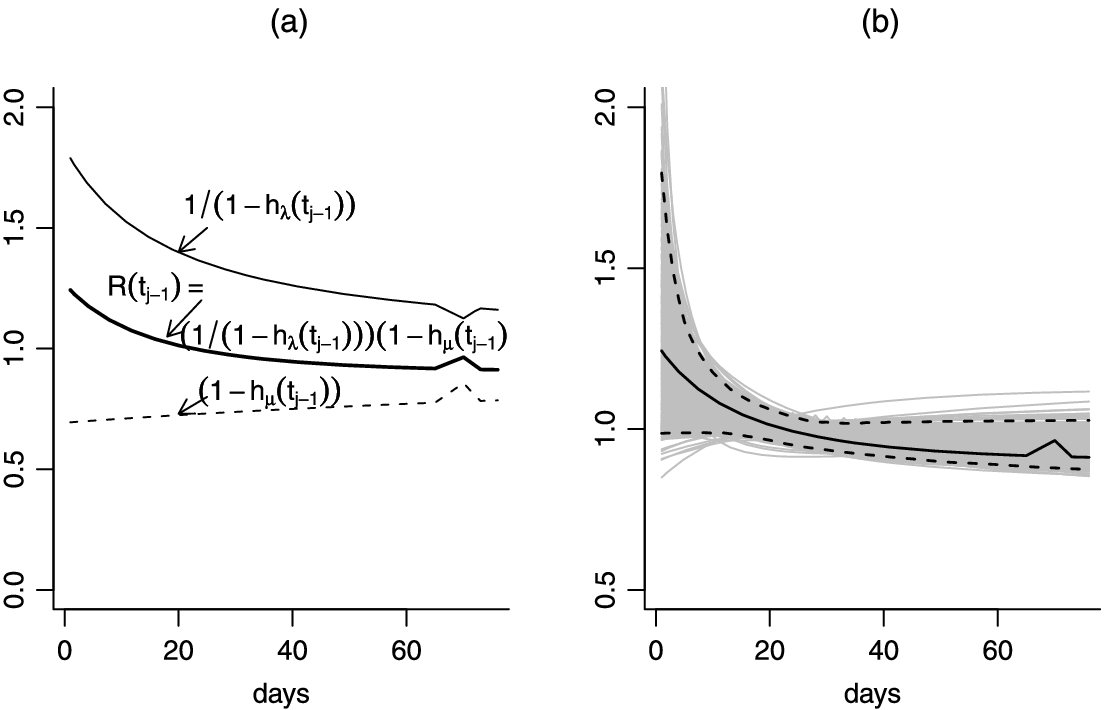}

\caption{Time dependency of birth and death rates
which jointly yield the effective reproduction number.
\textup{(a)} ($(1-h_{\mu}(t_{j-1}) $) indicates the rate at which an
infected-and-detected case escapes removal, whereas ($1/(1-h_{\lambda
}(t_{j-1})) $) denotes the rate at which a single infected-and-detected
(but not yet removed) case reproduces secondary cases. The product
$(1-h_{\mu}(t_{j-1})) $ $/(1-h_{\lambda}(t_{j-1}))$ yields the
effective reproduction number $R(t)$ as a function of time, which can
be interpreted as the average number of secondary cases generated by a
single primary case at time $t_{j-1}$. If $R(t_{j-1}) < 1$, it suggests
that the epidemic is in decline. In our example, the expected value of
$R(t_{j-1})$ declined below unity on day 23 since the detection of
index case. (b) The effective reproduction number, $R(t_{j-1})$,
calculated from sample paths drawn from the underlying estimated
process (gray lines) with the estimated value (black line) and the 95\%
percentile lines (dashed lines).}\label{fig3}
\end{figure}

Parameter estimates for the best fitting model are shown in
Table \ref{tab2} with their standard errors. The logarithm of the
acceleration factor, $d$, was estimated at $1.47$ with a standard error
of $0.369$, and the logarithm of the proportionality between the
reproductive power and the removal rate was $1.54$ with a standard
error of $0.355$. The estimates of mean reproduction and removal times
for the Lomax distribution can be calculated from [\citet{kleiber-kotz}]
$E(t)=\frac{b\Gamma(2)\Gamma(q-1)}{\Gamma(q)}$. The mean reproduction
time was estimated as $1.81$, indicating that it takes on average 1.8
days for a detected case to reproduce another detected case. The mean
removal time was estimated as $2.79,$ indicating that on average it
takes 2.8 days for a detected infected case to be removed.

In Figure \ref{fig3} the rate at which a single case survives
removal, ($ 1-h_{\mu}(t_{j-1}) $), is shown as a function of epidemic
date. In addition, the rate at which a single nonremoved case
reproduces secondary cases, ($1/(1-h_{\lambda}(t_{j-1}) $), is also
shown in the figure. The product of these two functions jointly yields
$R(t_{j-1})$ (also shown in Figure \ref{fig3}). Compared with other
modeling results [e.g., \citet{nishiura}], our estimates of $R(t_{j-1})$
are smoothed as a function of time owing to our parametric model for
the survival functions of reproduction and removal. Nevertheless, it
should be noted that our approach does not have to assume that the
generation time distribution is known [i.e., common assumption in
estimating $R(t)$], because our approach does not have to translate a
growth rate of incidence to the reproduction number. As we discussed
above, $R(t_{j-1}) < 1$ suggests that the epidemic is in decline at
time $t_{j-1}$ [vice versa, if $R(t_{j-1}) > 1$]. This can be
understood by considering the condition for $R(t_{j-1}) = 1$; that is,
the reproductive power becomes equivalent to the removal rate at time
$t$. In our example, the expected value of $R(t_{j-1})$ declined below
unity for the first time on day 23 since the detection of the index
case, supporting eventual end of the epidemic in the later stage. The
sawtooth at the end of the lines is considered to have been caused by
zero prevalence during the corresponding time period (i.e., because of
our conditional measurement, the survival functions reflect small
variations in the observed data). Figure \ref{fig3}(b) shows the
estimated effective reproduction number $R(t_{j-1})$. To get an idea of
the statistical uncertainty of $R(t_{j-1})$, 500 sample paths were
drawn from the estimated nonhomogeneous birth--death model and for each
sample path the effective reproduction number was calculated (gray
lines). The 95\% percentile lines of $R(t_{j-1})$ are also shown.
%

%
\section{Discussion}\label{sec6}
In the present study we modeled an epidemic based on the nonhomogeneous
birth--death process, addressing some of the critical issues which are
seen in the observation of directly transmitted infectious diseases.
First, we modeled infected-and-detected individuals, which corresponds
to observable and countable information in practice (e.g., our model
requires neither susceptibles nor infectious individuals). Second, for
a similar reason, the application of a birth--death process allowed the
population at risk (i.e., the susceptible population) to vary with
time. Third, applying the concepts of a nonhomogeneous birth--death
process to epidemic modeling, dependent events (i.e., dependence of a
single infected individual on other infected individuals) were
addressed in the model. Fourth, our stochastic model offered an
explicit likelihood function and yielded a standard error of
parameters. Last, our model allowed estimation of the effective
reproduction number, $R(t)$. Although a different probability
distribution was given by \citet{bailey}, the derivation was not given
in the literature, and, to the best of our knowledge, equation (\ref
{dbinom}) is the first to derive the pdf explicitly.\looseness=1

In a recent study \citet{vandenbroek} applied a nonhomogeneous
birth process to epidemic data, in which the survival-time distribution
was modified by a final-size parameter to describe the end of an
epidemic (which was influenced by public health countermeasures). The
countermeasures would not only reduce the final size of an epidemic but
also the reproductive power as a function of time, because secondary
transmissions caused by infected individuals are restricted under the
control measures. The nonhomogeneous birth process in the previous
study permitted an explicit assessment of the time variations in the
number of newly infected individuals (and thus the reproductive
power).\looseness=1

In the present study the proposed nonhomogeneous birth--death
process further improved our understanding of time dependency by
explicitly adding the nonhomogeneous removal rate. Whereas the
reproductive power changes as a function of time due to variations in
susceptible individuals or in the transmission rate, the removal rate
is also nonhomogeneous when infected individuals are likely to be
removed upon detection. Thus, adding nonhomogeneous death to a
nonhomogeneous birth process enabled us to separately consider the
effectiveness of countermeasures as a reduction in reproductive power
(e.g., reduction in infectious contacts) and an increase in removal
rate of infected individuals (e.g., culling of infected farms). In this
way, the fading out of an epidemic was modeled in a smoother way, as
compared to the previous model based on a nonhomogeneous birth process alone.

Moreover, it should be noted that our model does not
necessarily require a homogeneous mixing assumption to describe
contacts, because our assumptions of the time-dependent rates
implicitly include those nonhomogeneities. For example, our model
allows time variations in susceptible individuals. Nevertheless, our
model only accounts for the nonhomogeneity with respect to time in an
explicit manner, and understanding other heterogeneous aspects of
transmission requires further information.

Of course, there are many possible candidate distributions to model the
reproduction and the removal times. We have selected the Burr
distribution for three reasons:
\begin{enumerate}
\item As noted in Section \ref{sec4}, the Burr family coincides with our
analytical understanding of the epidemic modeling, especially at the
early stage of an epidemic.
\item Essentially, the Burr distribution is flexible and has special
and limiting cases. For instance, the distribution can be regarded as a
Weibull distribution with a random scale parameter.
\item In the context of the present study, the proportional hazard rate
and the accelerated event time interpretation may be very helpful in
model reduction and further interpretation of the data.
\end{enumerate}

As an example, the nonhomogeneous birth--death model was applied
to epidemic data of avian influenza A (H7N7) in the Netherlands, 2003,
showing that the model fitted to the data very well. Indeed, since the
data set has information on both birth and death events for each
individual case, the Dutch data appeared very useful for fitting
prevalence data and applying our modeling method. Even in the presence
of gaps in the right tail of the epidemic curve, and even though the
center was not well determined, our model reasonably described the
time-course of the observed epidemic. In particular, our model
permitted an estimation of the effective reproduction number $R(t)$ as
a function of time without imposing a specific distribution of the
generation time.

As is often the case with natural outbreaks, a single
observation represents just one sample path from the process for which
the above-mentioned model is imposed as the generator. There is no
random sampling of infectious disease outbreaks, and a repeated
sampling interpretation for the resulting model fit might be difficult.
In other words, the description and conclusions arising from analysis
of a single outbreak data set is valid only for that outbreak. To find
some general disease-specific conclusions from such an exercise, we
stress that it is important to analyze several different outbreaks for
the same disease. For such a purpose, one may use our model to
accumulate the experience of applying our method to several outbreaks.

\section*{Acknowledgments}
The authors greatly appreciate anonymous review comments which helped
improve this article.

\printaddresses

\end{document}